\documentclass[12pt]{article}
\usepackage{tikz}
\usetikzlibrary{matrix,arrows,positioning,shapes,snakes,fadings,decorations.pathmorphing,
decorations.pathreplacing,automata,shadows,decorations.markings,patterns}
\usepackage[thicklines]{cancel}
\usetikzlibrary{calc,shapes.callouts,shapes.arrows}
\usepackage{jheppub}
\usepackage{amsmath,amssymb,euscript,array,mathrsfs,appendix,ctable,marvosym}
\usepackage{arydshln}
\usepackage{todonotes}
\usepackage{graphicx}

\newcommand{\RE}{\operatorname{Re}}

\newcommand{\EQ}[1]{\begin{equation}\begin{split} #1
\end{split}\end{equation}}

\newcounter{Part}

\title{ Causality, Renormalizability  and Ultra-High Energy Gravitational Scattering}
\author{Timothy J. Hollowood and Graham M. Shore}
\affiliation{
Department of Physics, Swansea University,\\ Swansea, SA2 8PP, United Kingdom
}

\emailAdd{t.hollowood@swansea.ac.uk}
\abstract{The amplitude ${\cal A}(s,t)$ for ultra-high energy scattering can be found in the leading eikonal
approximation by considering propagation in an Aichelburg-Sexl gravitational shockwave background. 
Loop corrections in the QFT describing the scattered particles are encoded for energies below
the Planck scale in an effective action which in general exhibits causality violation and Shapiro time advances.
In this paper, we use Penrose limit techniques to calculate the full energy dependence of the
scattering phase shift $\Theta_{\rm scat}(\hat{s})$, where the single variable $\hat{s} = Gs/m^2 b^{d-2}$ 
contains both the CM energy $s$ and impact parameter $b$, for a range of scalar QFTs in $d$ dimensions
with different renormalizability properties. We evaluate the high-energy limit of  $\Theta_{\rm scat}(\hat{s})$
and show in detail how causality is related to the existence of a well-defined UV completion.
Similarities with graviton scattering and the corresponding resolution of causality violation in 
the effective action by string theory are briefly discussed.
}

\notoc
\begin{document}

\maketitle

\newpage

\section{Introduction}

At ultra-high energies, of order the Planck mass, the scattering of elementary particles is dominated by 
graviton exchange. Scattering at these energies is therefore an important theoretical laboratory in which 
to test fundamental ideas in quantum field theory, string theory and quantum gravity
(see refs.~\cite{'tHooft:1987rb,Muzinich:1987in,Amati:1987wq,Gross:1987kza,Gross:1987ar,Amati:1987uf,Amati:1988tn,
Amati:1990xe,Amati:1992zb,Amati:1993tb,Veneziano:2004er,Amati:2007ak,Giddings:2007bw,
Giddings:2009gj,Giddings:2011xs}).
The key technique used in the analysis is the eikonal approximation, which allows one to effectively 
sum up an infinite set of ladder diagrams for exchanging an arbitrary number of gravitons.

As a scattering problem, the natural quantity to define is the 
eikonal phase $\Theta(s,b)$, written in terms of the Mandelstam variable $s=4EE'$, where $E$ and $E'$ are the
energies of the particles, and the impact parameter $b$. 
The expression for the eikonal amplitude in terms of the usual pair of Mandelstam variables $(s,t)$ 
is then obtained by a Fourier transform,
\EQ{
{\cal A}(s,t=-\vec q^{\,2})=-2is\int d^{d-2}b\,e^{i\vec b\cdot\vec q}\left[e^{i\Theta(s,b)}-1\right]\ .
\label{a1}
}
In the ultra-high energy regime, we allow the dimensionless ratio $G s/b^{d-4}$ 
which determines the leading order eikonal phase to be large
(recall that in $d$ spacetime dimensions, $G = 1/M_p^{d-2}$).

As shown in \cite{'tHooft:1987rb}, the eikonal approximation allows the reformulation of the two-particle 
scattering problem in terms of the classical propagation of the first particle in the Aichelburg-Sexl shockwave
geometry \cite{Aichelburg:1970dh} produced by the other. The leading order phase $\Theta_{\rm cl}(s,b)$ 
is then given in terms of the discontinuous lightcone coordinate shift experienced by particle 1
as it passes the shockwave. For example, in four dimensions the phase shift is
$\Theta_{\rm cl} \sim - G s \log \left(b^2 \Lambda^2\right)$, for some cut-off $\Lambda$,
leading to the amplitude \cite{'tHooft:1987rb}
\EQ{
{\cal A}(s,t) = 8\pi i \,\frac{s}{t} \left(\frac{-t}{4\Lambda^2}\right)^{iGs} \frac{\Gamma(1-iGs)}{\Gamma(iGs)} 
\qquad  \implies \qquad
\left|{\cal A}(s,t)\right|^2 = (8\pi)^2\, \frac{G^2 s^4}{t^2} \ .
\label{a2}
}

Moving beyond this classical picture, in an interacting QFT the propagating particle has an associated vacuum polarization
cloud, characterised by the length scale $\lambda_c \sim 1/m$ of the virtual particles in the loop. In the shockwave
background, this is subject to gravitational tidal forces, which give a new quantum contribution to the phase shift,
\EQ{
\Theta(s,b) = \Theta_{\rm cl}(s,b) + \Theta_{\rm scat}(\hat{s}) \ ,
\label{a3}
}
where, as we see later, this further shift depends only on the combination
$\hat{s} = Gs/m^2 b^{d-2}$ of the CM energy and impact parameter.

At low energies, $\Theta_{\rm scat}(\hat{s})$ is described by an effective action describing the coupling of the curvature
to the quantum fields. This coupling violates the strong equivalence principle and is known in many instances to 
produce apparent causality violations in the form of superluminal propagation. In the context of gravitational shockwave
scattering, this is manifested as a Shapiro {\it time advance}. However, in order to determine
whether or not such a potential causality violation is really physical, we have to look at the high-energy limit. 
This means going beyond the effective theory and determining the scattering phase $\Theta_{\rm scat}(\hat{s})$
in the full, UV complete QFT.

In a recent paper \cite{Hollowood:2015elj}, we have shown by explicit calculation of the full energy dependence
of the scattering phase, how in a renormalisable theory, QED in four dimensions, 
the apparent causality violations arising from $\Theta_{\rm scat}(\hat{s})$ in the effective theory
are resolved in the UV limit. It is an interesting question, which we leave for future work, how these apparent
causality problems manifest themselves in the scattering amplitude ${\cal A}(s,t)$, how they are related to 
the unitarity properties of ${\cal A}(s,t)$ and the associated partial wave amplitudes, and how they are
resolved by the $\hat{s}\rightarrow\infty$ limit in the QFT picture of propagation in the shockwave spacetime.

Here, as a prelude to such investigations, we study the IR and UV properties of the scattering phase
$\Theta_{\rm scat}(\hat{s})$ for a range of QFTs exhibiting different renormalizability properties
in order to gain a clearer understanding of the interplay of causality, unitarity and renormalizability in the
presence of the shockwave background. 
This was prompted by the observation in \cite{Hollowood:2015elj} that the UV behaviour of the 
scattering phase in a purely scalar, super-renormalizable analogue of QED was quite different from QED itself, 
although still maintaining causality. To this end, in this paper we study the scattering problem for a class
of self-interacting $\phi^n$ scalar theories in $d$ dimensions for arbitrary $n$ and $d$,
and investigate in detail the relation of causality and renormalizability in the gravitational shockwave
background. Quite generally, the study of effective field theories in gravitational
shockwave spacetimes is a rigorous test of what constraints are placed on the form and values of the 
couplings in an effective theory in order that it admits a consistent UV completion.

There is a clear parallel between this programme and the pure gravity case, where the scattering particles are
themselves gravitons. In the paper \cite{Camanho:2014apa}, the effective field theory is taken to be the 
Einstein action augmented by a Gauss-Bonnet term.
As shown there, this effective theory exhibits superluminal causality violation. In this case, one resolution is that the 
effective theory must be embedded in a UV complete theory containing an infinite set of higher spin states 
that Reggeizes the amplitude as in string theory \cite{Camanho:2014apa,D'Appollonio:2015gpa}.
Note here the crucial r\^ole played by the introduction of the string scale $\lambda_s$, which is analogous
to the scale $\lambda_c$ in our QFT problem. Another interesting issue relevant to the Gauss-Bonnet case, is whether the causality-violating configuration of gravitational shockwaves can actually be engineered in the first place \cite{Papallo:2015rna}.

\subsection{Ultra-high energy scattering and shockwaves}

As  discussed above, ultra-high energy scattering can be viewed in the eikonal approximation in terms 
of propagation in a gravitational shockwave background, as illustrated in fig.~1.
\begin{figure}[h]
\begin{center}
\begin{tikzpicture}[scale=0.7,fill=black!20,decoration={markings,mark=at position 0.5 with {\arrow{>}}}]
\begin{scope}[xshift=15.5cm,yshift=0.4cm]
\node at (0.8,0.2) {$=$};
\end{scope}
\begin{scope}[xshift=9cm,yshift=0cm]
\begin{scope}[scale=0.9]
\draw[thick,decoration={aspect=0.6, segment length=1.5mm, amplitude=0.8mm,coil},decorate] (0,1.5) -- (0,0);
\draw[thick,decoration={aspect=0.6, segment length=1.5mm, amplitude=0.8mm,coil},decorate] (0.5,1.5) -- (2,0);
\draw[thick,decoration={aspect=0.6, segment length=1.5mm, amplitude=0.8mm,coil},decorate] (2,1.5) -- (0.5,0);
\draw[thick,decoration={aspect=0.6, segment length=1.5mm, amplitude=0.8mm,coil},decorate] (2.5,1.5) -- (2.5,0);
\draw[thick,decoration={aspect=0.6, segment length=1.5mm, amplitude=0.8mm,coil},decorate] (3,1.5) -- (3.5,0);
\draw[thick,decoration={aspect=0.6, segment length=1.5mm, amplitude=0.8mm,coil},decorate] (5,1.5) -- (4.5,0);
\draw[thick,decoration={aspect=0.6, segment length=1.5mm, amplitude=0.8mm,coil},decorate] (5.5,1.5) -- (5.5,0);
\draw[very thick] (-1,2.5) -- (0,1.5) -- (3.5,1.5);
\draw[very thick,densely dashed] (3.5,1.5) -- (4.5,1.5);
\draw[very thick]  (4.5,1.5) -- (5.5,1.5) -- (6.5,2.5);
\draw[very thick] (-1,-1) -- (0,0) -- (3.5,0);
\draw[very thick,densely dashed] (3.5,0) -- (4.5,0);
\draw[dashed] (3.5,0.75) -- (4.5,0.75);
\draw[very thick] (4.5,0) -- (5.5,0) -- (6.5,-1);
\end{scope}
\node at (-2,0.75) {$\Large\displaystyle\sum$};
\node at (-2,-0.2) {\footnotesize gravitons};
\end{scope}
\begin{scope}[xshift=18cm,yshift=-1.2cm,scale=0.6]
\draw[line width=2mm,color=black!20] (7,0) --  (0,7);
\draw[very thick,postaction={decorate}] (0,0) -- (3.5,3.5);
\draw[very thick,postaction={decorate}] (2.5,4.5) -- (5,7);
\draw[very thick,postaction={decorate}] (3.5,3.5) -- (2.5,4.5);
\node[rotate=45] at (1,2) {$v=0$};
\node[rotate=-45] at (6.4,1.8) {$u=0$};
\node at (6.8,7.7) {$v=\Delta v$}; \end{scope}
\end{tikzpicture}
\caption {\footnotesize  The tree level gravitational scattering of two high energy particles can be improved by 
summing over ladder diagrams (including crossings) involving multiple graviton exchange. 
This re-summation is equivalent to studying the classical motion, the null geodesics, of one of the particles 
in the Aichelburg-Sexl shockwave produced by the other. 
The scattering is then described by a Shapiro time delay, or advance, of the particle as it crosses the shockwave.}
\label{f1}
\end{center}
\end{figure}
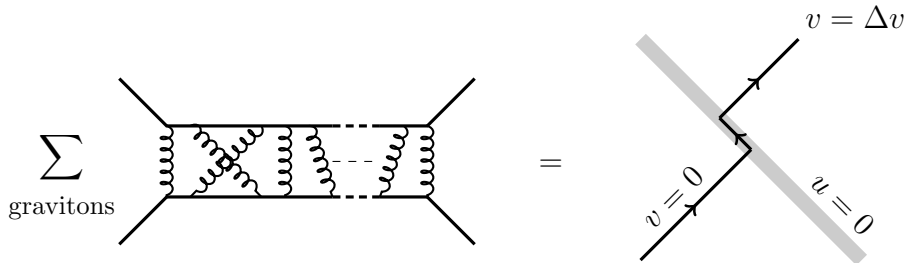

A central role is played, therefore, by the Aichelburg-Sexl metric \cite{Aichelburg:1970dh},
\begin{equation}
ds^2 = -2 du \,dv + f(r) \delta(u) du^2 +\sum_{i=1}^{d-2} dx^i\,dx^i\ ,\qquad r^2=\sum_{i=1}^{d-2}x^ix^i\ .
\label{ba}
\end{equation}
The profile function $f(r)$ is determined by the Ricci curvature $R_{uu} = 8\pi G T_{uu}$
and depends on the nature of the matter source for the shockwave. 
For the case of an ultra-high energy particle of energy $E'$, the profile function $f(r)$ is
\EQ{
f(r)=\frac{4\Gamma(\frac{d-4}2)}{\pi^{\frac{d-4}2}}\cdot\frac{GE'}{r^{d-4}}\ .
\label{bb}
}
This particle follows the trajectory $u=x^i=0$, {\it i.e.}~$r=0$.

At very high energy, i.e.~for the scalar field $E\gg m$, particle 1 then follows a null geodesic propagating 
in the opposite direction to the shockwave, that is $v=0$ and impact parameter $r=b$. The fact that we 
can talk about particle trajectories means that we are working in the geometric optics limit, 
which requires $E\gg\sigma$, where $\sigma$ is the curvature scale expressed as a mass scale. 
For the shockwave, $\sigma\sim GE'/b^{d-2}$.

Along the geodesic followed by particle 1, the null coordinate takes the form
\begin{equation}
v = \frac{1}{2} f(b) \vartheta(u) + \frac{1}{8} f'(b)^2 u \vartheta(u) \ .
\label{ab}
\end{equation}
The first term here, corresponds to a instantaneous jump in the null coordinate $v$ as the particle 
passes through the shockwave wavefront at $u=0$: 
\EQ{
\Delta v=\frac12f(b)=\frac{2\Gamma(\frac{d-4}2)}{\pi^{\frac{d-4}2}}\cdot\frac{GE'}{b^{d-4}}\ ,
}
In four dimensions, this is
\EQ{
(d=4):\qquad \Delta v= -2G E'\log\Big(\frac b{r_0}\Big)^2 , 
}
where $r_0 = 1/\Lambda$ is short-distance regulator.\footnote{This can be established by regularizing the 
particle as a beam shockwave \cite{Ferrari:1988cc} with radius $r_0$.} In $d=4$, therefore, 
the shift in the null coordinate is actually a Shapiro time advance. 

Thinking of a wave packet with a narrow spread of energies implies that the eikonal phase itself
is related to the shift in the null coordinate by
\EQ{
\Delta v=\RE\frac{\partial\Theta}{\partial E}\ .
\label{zz1}
}
Note that, in general, the eikonal phase can have a imaginary part which is interpreted as a 
modulation of the amplitide of the mode. In the shockwave case, therefore,
\EQ{
\Theta_{\rm cl}(s,b)=\frac{\Gamma(\frac{d-4}2)}{2\pi^{\frac{d-4}2}}\cdot\frac{Gs}{b^{d-4}}\ .
}

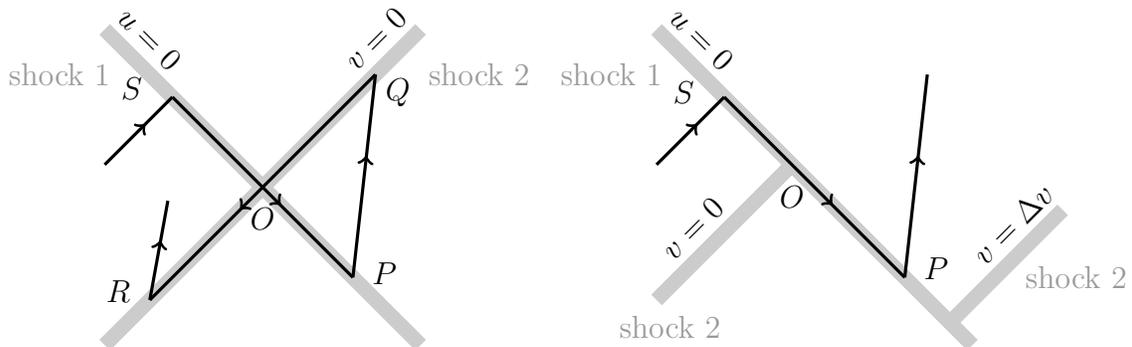
\begin{figure}[h]
\begin{center}
\begin{tikzpicture}[scale=0.6,fill=black!20,decoration={markings,mark=at position 0.6 with {\arrow{>}}}]
\draw[line width=2mm,color=black!20] (0,0) -- (7,7);
\draw[line width=2mm,color=black!20] (7,0) --  (0,7);
\draw[very thick,postaction={decorate}] (0,4) -- (1.5,5.5);
\draw[very thick,postaction={decorate}] (1.5,5.5) -- (5.5,1.5);
\draw[very thick,postaction={decorate}] (5.5,1.5) -- (6,6);
\draw[very thick,postaction={decorate}] (6,6) -- (1,1);
\draw[very thick,postaction={decorate}] (1,1) -- (1.4,3.2);
\node at (0.6,5.7) {$S$};
\node at (6.2,1.6) {$P$};
\node at (6.5,5.6) {$Q$};
\node at (0.3,1.2) {$R$};
\node at (3.5,2.8) {$O$};
\node[rotate=45] at (6,6.8) {$v=0$};
\node[rotate=-45] at (1,6.8) {$u=0$};
\node[opacity=0.4] at (-1,6) {shock 1};
\node[opacity=0.4] at (8.3,6) {shock 2};
\end{tikzpicture}
\begin{tikzpicture}[scale=0.6,fill=black!20,decoration={markings,mark=at position 0.6 with {\arrow{>}}}]
\draw[line width=2mm,color=black!20] (0,1) -- (3,4);
\draw[line width=2mm,color=black!20] (6.5,0.5) -- (9,3);
\draw[line width=2mm,color=black!20] (7,0) -- (0,7);
\draw[very thick,postaction={decorate}] (0,4) -- (1.5,5.5);
\draw[very thick,postaction={decorate}] (1.5,5.5) -- (5.5,1.5);
\draw[very thick,postaction={decorate}] (5.5,1.5) -- (6,6);
\node at (0.6,5.6) {$S$};
\node at (6.2,1.7) {$P$};
\node at (3,3.3) {$O$};
\node[rotate=45] at (0.8,2.6) {$v=0$};
\node[rotate=45] at (7.9,2.7) {$v=\Delta v$};
\node[rotate=-45] at (1,6.8) {$u=0$};
\node[opacity=0.4] at (-1,6) {shock 1};
\node[opacity=0.4] at (9.3,1.5) {shock 2};
\node[opacity=0.4] at (0.3,0.4) {shock 2};
\end{tikzpicture}
\caption{\footnotesize Left: the proposed time machine consisting of two shockwaves 
moving in opposite directions that collide at $O$. The particle collides with the first at $S$, 
experiences a shift back to $P$ which then allows it to catch up with shockwave 2 with a 
jump back to $R$ in the past lightcone of $S$. Right: in the true picture, the wavefront of 
shockwave 2 at the same impact parameter as the particle undergoes the same shift 
$\Delta v<0$ as the particle. It is clear that the particle can, therefore never catch up 
with the shockwave 2 to complete the circuit shown on the left.}
\label{f2}
\end{center}
\end{figure}

The time advance in the case $d=4$ is interesting: does this imply a breakdown of causality? 
The answer turns out to be quite subtle. One way to approach it is to construct a geometry, 
a ``time machine", that allows particles to propagate around a closed trajectory. 
The simplest setup, first introduced in \cite{Shore:2002in} and then considered in 
\cite{Adams:2006sv,Camanho:2014apa}, consists of two shockwaves that collide wth some 
impact parameter $L$: see the left-hand side of fig.~\ref{f2}.

In fact, a careful analysis \cite{Shore:2002in,Hollowood:2015elj} shows that the time machine fails to work. 
Essentially, the equivalence principle means that one shockwave jumps back in the background 
of the other shockwave just before the particle jumps back. So the particle can never catch up 
with the second shockwave: see the right-hand side of fig.~\ref{f2}.

Another way to think about this is that geometrically the classical time delay or advance is a coordinate-dependent
effect, which may be removed by working in Rosen-like coordinates where the particle trajectory is continuous
across the shock. The price to be paid is that in these new coordinates, the regions behind the shockwaves
are no longer manifestly flat. In order to assign a physical significance to the classical shift in the context of
scattering, we therefore need to impose some external identification of the asymptotically past and future 
Minkowski space regions.

\subsection{Curvature couplings and the effective theory}

To motivate our discussion, consider first the case of graviton scattering. Here, the effective action
considered by \cite{Camanho:2014apa} containing the Gauss-Bonnet term is 
\EQ{
S_\text{eff}=\frac1{16\pi G}\int d^dx\,\sqrt{g}\big[R+\alpha\big(R^2_{\mu\nu\rho\sigma}-
4R_{\mu\nu}^2+R^2\big)+\cdots\big]\ ,
}
which is non-trivial when $d\neq4$.
The Gauss-Bonnet coupling in the gravitational effective action violates the strong equivalence principle (SEP) 
and the scattered graviton 1 no longer propagates along null geodesics. It induces a Shapiro time delay or advance 
of the form  \cite{Camanho:2014apa}
\EQ{
\Delta v\sim\pm \alpha\frac{GE'}{b^{d-2}}\ ,
}
where the sign depends on the graviton polarization. In this case, the time delay or advance is a 
genuine non-coordinate dependent effect that can be used to set up a causality paradox by using 
the two-shockwave time machine described above.

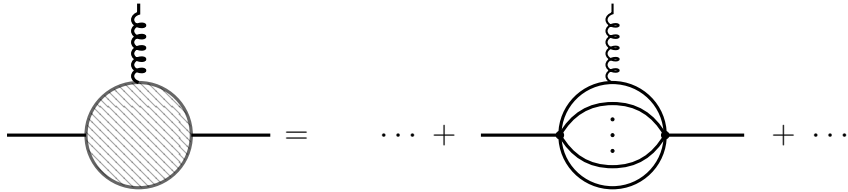
\begin{figure}[h]
\begin{center}
\begin{tikzpicture}[scale=0.7,fill=black!20,decoration={markings,mark=at position 0.6 with {\arrow{>}}}]
\begin{scope}[xshift=0cm]
\draw[very thick] (-1.5,0) -- (0,0);
\draw[very thick] (2,0) -- (3.5,0);
\fill[very thick,pattern=north west lines,opacity=.6,draw] (1,0) circle [radius=1];
\draw[very thick,decoration={aspect=0.6, segment length=1.5mm, amplitude=0.8mm,coil},decorate] (1,1) -- (1,2.5); 
\end{scope}
\begin{scope}[xshift=9cm]
\node at (4.8,0) {$+\ \cdots$};
\node at (-3.6,0) {$=\qquad\cdots\ +$};
\draw[very thick] (-1.5,0) -- (0,0);
\draw[very thick] (2,0) -- (3.5,0);
\draw[very thick] (0,0) to[out=90,in=-180] (1,1) to[out=0,in=90] (2,0);
\draw[very thick] (0,0) to[out=-90,in=180] (1,-1) to[out=0,in=-90] (2,0);
\draw[very thick] (0,0) to[out=60,in=-180] (1,0.6) to[out=0,in=120] (2,0);
\draw[very thick] (0,0) to[out=-60,in=180] (1,-0.6) to[out=0,in=-120] (2,0);
\filldraw[black] (0,0) circle (2pt);
\filldraw[black] (2,0) circle (2pt);
\filldraw[black] (1,0.3) circle (0.8pt);
\filldraw[black] (1,-0.3) circle (0.8pt);\filldraw[black] (1,0) circle (0.8pt);
\draw[thick,decoration={aspect=0.6, segment length=1.5mm, amplitude=0.8mm,coil},decorate] (1,1) -- (1,2.5);
\end{scope}
\end{tikzpicture}
\caption{\footnotesize The curvature-dependent coupling in the effective theory arises from one-graviton 
corrections to the self energy, which at leading order in perturbation theory involve the diagram shown 
where the graviton couples to one of the particles in the loops.}
\label{f3}
\end{center}
\end{figure}

This is analogous to the effective action generated in a self-interacting scalar QFT through the
coupling of a background graviton to the self-energy loop of the propagating scalar particle.
This induces a SEP-violating coupling to the Ricci tensor:\footnote{This is the scalar field equivalent of
the original Drummond-Hathrell effective action for QED in a background gravitational field, where
the effect of gravitational tidal forces inducing superluminal phase velocities at low frequency
was first discovered \cite{Drummond:1979pp}.}
\begin{equation}
S_\text{eff}= \int d^d x\,\sqrt{g} \left[-\frac{1}{2}\partial^\mu \phi \partial_\mu \phi -
\frac{m^2}2\phi^2-\frac{\lambda}{n!}\phi^n-\alpha R^{\mu\nu}\partial_\mu \phi \partial_\nu \phi +\cdots\right]\ .
\label{cf1}
\end{equation}
Since in general the couplings $\lambda$ are dimensionful, it is convenient to define the
dimensionless coupling $\tilde{\lambda}$ as
\EQ{
\tilde\lambda=m^{2-p}\lambda ,  \qquad\qquad   p=\frac{(n-2)(d-2)}2\ . 
\label{dimp}
}
The leading order contribution in perturbation theory is shown in fig.~\ref{f3} and leads to a coupling
\EQ{
\alpha=c_1\frac{\tilde\lambda^2}{m^2}\ ,
}
where $c_1$ is a {\it positive\/} number.

Unlike the gravitational case, there is no polarization dependence here.
Nevertheless, for a background that satisfies the null energy condition, the coupling leads to a Shapiro time advance. 
Note also that the shockwave produced by a high energy particle is a vacuum solution of Einstein's equation 
and so is Ricci flat. However, one can consider the  related ``beam shockwave" \cite{Ferrari:1988cc}, 
described in detail in section \ref{s3}, that does have a non-vanishing Ricci curvature. 
In this case one finds a time advance,
\EQ{
\Delta v=-c_2\tilde\lambda^2\frac{G\mu}{m^2}\ ,
\label{yy1}
}
where $\mu$ is the energy density of the beam and $c_2$ is a positive constant. 
The particle shockwave produces a time advance at higher order in the curvature expansion.

The time shift is actually a completely generic effect for propagation in any non-Ricci flat background
as can be seen by solving the linearized equation of motion. Working
in the eikonal, or geometric optics, limit $E\gg m\gg\sigma$,
where $\sigma$ is the curvature scale of the shockwave expressed as an energy scale,
the solution for the field takes the form of a rapidly varying phase:
\EQ{
\phi(x)\thicksim\exp(-i\Theta(x))\ .
}
In this limit, the phase $\Theta(x)$ defines a congruence of null geodesics, corresponding to the rays 
of geometric optics, whose tangent vector field 
is $\partial^\mu\Theta$. It is always possible to introduce a set of adapted coordinates 
$(u,V,X^i)$, $i=1,2,\ldots,d-2$, so that the congruence is described by 
$V=\text{const.}$ and $X^i=\text{const.}$ and for which 
\EQ{
\Theta(x)=EV\ .
}

Now consider the modifications implied by the curvature coupling in \eqref{cf1}. 
Working in perturbation theory, we find that the phase receives an additional contribution
\EQ{
\Theta(x)=EV-\frac{\alpha E}2\int^u du\,R_{uu}(u)\ ,
}
where $R_{uu}(u)$ is a component of the Ricci tensor evaluated along the null geodesic $V=X^i=0$. 
If the curved region is concentrated in an interval $[u_1,u_2]$, then the effect of the coupling is 
to introduce a Shapiro time advance,
\EQ{
\Delta v=-\frac{\alpha}2\int_{u_1}^{u_2}du\,R_{uu}(u)\ .
}
The fact that this is an advance, {\it i.e.}~negative, is because the null energy condition
 implies that $R_{uu}\geq0$. It is also a fact, that we establish later, that $\alpha>0$.

Another way to think of this result is to notice that in the effective action \eqref{cf1}, it is as if the 
field propagates in an effective metric ${\cal G}_{\mu\nu}=g_{\mu\nu}+\alpha R_{\mu\nu}$. 
The particle's wave-vector $k_\mu=\partial_\mu\Theta$ is null with respect to this effective metric 
but {\it spacelike\/} with respect to the real metric. This corresponds to superluminal propagation 
and a Shapiro time advance.

\subsection{Beyond the effective action}

Our central theme is to address the question, in the context of scalar fields, of how these 
curvature couplings impact on causality. A causality violating effect seen in the low energy theory is not, 
in itself, a problem because such issues should be addressed in the UV limit. 
The intuition here is that in order to send information from one place to another, it is necessary 
to use sharp-fronted wave packets that inevitably require high frequency (energy) modes: these issue 
are discussed at length in \cite{Shore:2007um}. 
The question then is exactly how high does the energy have to be. 
The point is that if the Shapiro time advance \eqref{yy1} is to be observable effect, we need that
\EQ{
E \Delta v>1\ .
}
If $\sigma$ is the curvature scale, which is proportional to $GE'/b^{d-2}$ for the particle shockwave, 
then this requires $\tilde\lambda^2 E \sigma/m^2>1$. 
This means that in order to access the UV limit necessary to discuss causality, we need
to be able to work in the regime where  $\tilde{\lambda}^2 \hat{s} > 1$,
where $\hat{s} = Gs/m^2 b^{d-2}$ is the dimensionless variable in the scattering phase introduced earlier.

\begin{figure}[h]
\begin{center}
\begin{tikzpicture}[scale=0.8,fill=black!20,decoration={markings,mark=at position 0.6 with {\arrow{>}}}]
\node at (-2.5,-0.8) {\footnotesize gravitons};
\node at (-2.5,0) {$\Large\displaystyle\sum$};
\draw[very thick] (-1.5,0) -- (0,0);
\draw[very thick] (2,0) -- (3.5,0);
\draw[very thick] (0,0) to[out=90,in=-180] (1,1) to[out=0,in=90] (2,0);
\draw[very thick] (0,0) to[out=-90,in=180] (1,-1) to[out=0,in=-90] (2,0);
\draw[very thick] (0,0) to[out=60,in=-180] (1,0.6) to[out=0,in=120] (2,0);
\draw[very thick] (0,0) to[out=-60,in=180] (1,-0.6) to[out=0,in=-120] (2,0);
\filldraw[black] (0,0) circle (2pt);
\filldraw[black] (2,0) circle (2pt);
\filldraw[black] (1,0.3) circle (0.8pt);
\filldraw[black] (1,-0.3) circle (0.8pt);
\filldraw[black] (1,0) circle (0.8pt);
\draw[thick,decoration={aspect=0.6, segment length=1.5mm, amplitude=0.8mm,coil},decorate] (1,1) -- (1,2);
\draw[thick,decoration={aspect=0.6, segment length=1.5mm, amplitude=0.8mm,coil},decorate] (0.5,0.5) -- (0,1.7);
\draw[thick,decoration={aspect=0.6, segment length=1.5mm, amplitude=0.8mm,coil},decorate] (1.5,-0.5) -- (2,-1.7);
\draw[thick,decoration={aspect=0.6, segment length=1.5mm, amplitude=0.8mm,coil},decorate] (0.5,-0.5) -- (0,-1.7);
\draw[thick,decoration={aspect=0.6, segment length=1.5mm, amplitude=0.8mm,coil},decorate] (2.5,0) -- (2.8,1.3);
\begin{scope}[scale=0.6,xshift=19cm,yshift=-3cm]
\node at (-10,3) {$=$};
\draw[line width=2mm,color=black!20] (-1,0) -- (-7,6);
\draw[very thick] (-7,0) -- (-5.5,1.5); 
\draw[very thick]  (-2.5,4) -- (-1.5,6.4); 
\draw[very thick,] (-5.5,1.5) to[out=30,in=-170] (-3.7,3.3);
\draw[very thick,] (-5.5,1.5) to[out=-40,in=-150] (-3.7,1.2);
\draw[very thick,] (-5.5,1.5) to[out=60,in=170] (-3.7,3.7);
\draw[very thick,] (-5.5,1.5) to[out=-80,in=-160] (-3.7,0.8);
\draw[very thick] (-3.7,3.3) to[out=10,in=160] (-2.5,4);
\draw[very thick] (-3.7,1.2) to[out=30,in=-120] (-2.5,4);
\draw[very thick] (-3.7,3.7) to[out=10,in=100] (-2.5,4);
\draw[very thick] (-3.7,0.8) to[out=30,in=-80] (-2.5,4);
\filldraw[black] (-5.5,1.5) circle (4pt);
\filldraw[black] (-2.5,4) circle (4pt);
\filldraw[black] (-4.3,2.6) circle (1pt);
\filldraw[black] (-4,2.1) circle (1pt);
\filldraw[black] (-3.7,1.6) circle (1pt);
\end{scope}
\end{tikzpicture}
\caption{\footnotesize The eikonal approximation can be extended by summing over diagrams 
where an arbitrary number of gravitons are attached to the order $\lambda^2$ contribution to the self energy. 
This is equivalent to evaluating the original diagram with propagators defined in the curved spacetime 
as illustrated schematically in the right-hand figure.}
\label{f4}
\end{center}
\end{figure}
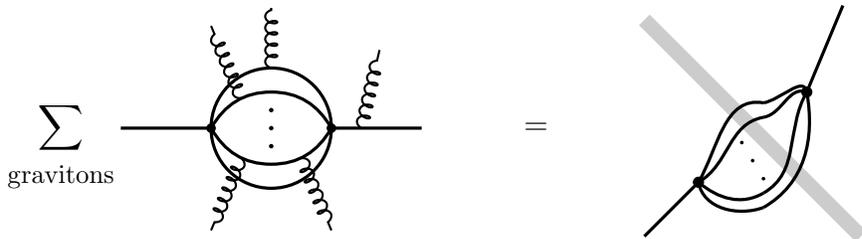

Our main result will be to extend the determination of the scattering phase $\Theta_{\rm scat}(\hat{s})$
and Shapiro time advance
from the low-energy effective action to the UV limit where we can resolve issues with causality.
This calculation may be thought of as summing over the leading order self-energy diagram with an 
arbitrary number of external gravitons attached as shown in fig.~\ref{f4}, which is equivalent to 
calculating the original diagram with propagators defined in the curved shockwave geometry.

We are able to evaluate the self-energy diagram with curved space propagators given by the shockwave 
geometry and determine the complete $\hat{s}$-dependence of the scattering phase.
Our results are valid for $E \gg m \gg \sigma$, where $\sigma\sim GE'/b^{d-2}$ for the particle shockwave
and $\sigma\sim G\mu$ for the beam shockwave.
We give an expression for the complete $\hat{s}$-dependence of $\Theta_{\rm scat}(\hat{s})$, focusing on the
low-energy regime $\hat{s} \ll 1$ which reproduces the effective action result, and on the high-energy,
regime where $\hat{s} \gg 1$. Notice that since $\hat{s}$ involves both the energy and impact parameter,
our calculation also allows us to access the small-$b$ scattering regime.

Self-interacting $\phi^n$ scalar field theories in arbitrary dimensions therefore provide a nice testing ground 
for studying the effect of renormalizability on the realisation of causality in gravitational shockwave backgrounds, 
complementing the discussion of QED in four dimensions in our recent paper \cite{Hollowood:2015elj}. 
We will show that when a UV completion exists, that is when the spacetime dimension is at, or below, 
the critical dimension $d\leq d_\text{crit}= 2n/(n-2)$ (equivalent to $p=2$ in (\ref{dimp})), 
{\it i.e.}~in the renormalizable case, 
the high energy limit of the Shapiro time advance goes to zero; causality is respected and a time machine 
cannot be created. On the other hand, if $d>d_\text{crit}$, {\it i.e.}~the theory is non-renormalizable, 
the Shapiro time advance persists and indeed diverges at the lowest order in perturbation theory.  
(Of course, this does not necessarily rule out the possibility that causality could be repaired 
at higher orders.) Within these broad categories, we will find differences in the $\hat{s}$-dependence
of the theories with various $n$ and $d$, reflecting power counting. These asymptotic behaviours 
for the phase $\Theta_{\rm scat}(\hat{s})$ naturally also determine the scattering amplitude ${\cal A}(s,t)$
itself.

\section{The Eikonal Phase}\label{s2}

In this section, we calculate the contribution to the eikonal phase from the curved background 
to leading order in perturbation theory (the calculation is similar to the analysis of QED 
in \cite{Hollowood:2007kt,Hollowood:2007ku,Hollowood:2008kq,Hollowood:2009qz,
Hollowood:2010bd,Hollowood:2011yh}). 

The task involves solving the linearized quantum corrected equation of motion.
\EQ{
\frac1{\sqrt{g}}\partial_\mu\Big(\sqrt{g}\,\partial^\mu\phi(x)\Big)-m^2\phi(x)=
\int d^dx\,\sqrt{g'}\,\Pi_\text{ret}(x,x')\phi(x')\ ,
\label{aa1}
}
in an appropriate approximation scheme. Note that $\Pi_\text{ret}(x,x')$ is the retarded 
(Schwinger-Keldysh) self-energy or vacuum polarization calculated in the curved space background. 
In order to solve \eqref{aa1}, it is necessary to define appropriate boundary conditions. This will be 
described in more detail later.

To start with, we work in perturbation theory, beginning with a solution that describes a highly 
relativistic particle with energy $E\gg m$ propagating along a null geodesic $\gamma$, or more
precisely a null congruence containing $\gamma$. 
Associated with the congruence are a set of adapted coordinates $(u,V,X^i)$, $i=1,2,\ldots,d-2$, 
such that $\gamma$ corresponds to $V=X^i=0$ and $u$ is the affine coordinate.

As well as working in perturbation theory, we make the additional approximation that 
the mass of the field is much larger than the curvature scale $m\gg\mathfrak R$ 
(expressed as a mass scale) transverse to $\gamma$. The implication is that it is possible to approximate 
the background geometry with that in a tubular neighbourhood of $\gamma$. 
This is precisely the Penrose limit \cite{Penrose} of the metric associated to $\gamma$, which in terms 
of the adapted coordinates takes the form
\begin{equation}
ds^2\Big|_{\text{Penrose}} =-2 du \,dV + C_{ij}(u) dX^i\, dX^j\ .
\label{bg}
\end{equation}

Once the null geodesic $\gamma$ has been picked out, the problem is to solve \eqref{aa1} in the 
Penrose limit geometry in perturbation theory. The geometry \eqref{bg} is a plane wave and such 
spacetimes are WKB, or eikonal, exact. This means that the solution of the wave equation and the 
free Green functions are known exactly. Firstly, the solution of the wave equation takes the form
\EQ{
\phi(x) =g(u)^{-1/4}\exp\Big[-ip_-V -\frac i2p_+u+ ip_i X^i + \frac i{2E}p_i\psi^{ij}(u)p_j\Big] \ ,
\label{cb}
}
where
\EQ{
\psi^{ij}(u) = \int^udu\,[C^{-1}(u)]^{ij}\ .
\label{cc}
}
We will choose the solution with no transverse momentum $p_i=0$ and for an ultra-relativistic particle 
$p_-=E\gg m$ and $p_+=m^2/E$:
\EQ{
\phi(x)=g(u)^{-1/4}\exp\Big[-iEV -\frac{im^2}{2E}u\Big]\ .
}
The idea, working in the eikonal approximation, is to search for a  solution of the quantum corrected 
equation of motion \eqref{aa1} in the form
\EQ{
\phi(x)=g(u)^{-1/4}\exp\Big[-iEV -\frac{im^2}{2E}u+ i\Theta(u)\Big]\ .
}
Using the fact that the right-hand side is perturbatively small gives us the following equation for the 
eikonal phase,
\EQ{
\partial_u\Theta(u)=\frac2E\int du'\,dV'\,d^{d-2}X'\,\big[g(u)g(u')\big]^{1/4}\, 
\Pi(u,0;x')\exp\Big[\frac{im^2}{2E}(u-u')-iEV'\Big]
\label{ab2}
}

Now we consider the perturbative expansion of the vacuum polarization. There are various contributions 
at order $\lambda$. These are independent of curvature and correspond to mass renormalization or are 
cancelled by counter terms. The first important curvature-dependent contribution appears at order 
$\lambda^2$: see fig.~\ref{f3} and fig.~\ref{f4}. We will limit ourselves to calculating the curvature 
dependence of this diagram.

The contribution to the vacuum polarization from this diagram, in real space, is 
simply\footnote{There are other contributions that serve to renormalize the operators $\phi^k$, $k<n$, 
that do not affect the eikonal phase.} 
\EQ{
\Pi(x,x')=\lambda^2 G(x,x')^{n-1}+\cdots\ ,
}
where $G(x,x')$ is the free propagator in the plane-wave background. One important point is that, 
since we are integrating this against a positive energy mode, as in \eqref{ab2}, this automatically picks out 
the retarded component of the vacuum polarization since the integral is only non-vanishing when $u'\leq u$.

In a plane-wave geometry, the Green function is known exactly,
\EQ{
G(x,x')=\sqrt{\Delta(x,x')}\int_0^\infty\frac{dT}{(4\pi iT)^{d/2}}\,i\exp\Big[-im^2T+\frac{\sigma(x,x')}{2iT}\Big]\ ,
\label{db}
}
where, in the Rosen coordinates $(u,V,X^i)$, the geodesic interval is
\EQ{
\sigma(x,x')=-(u-u')(V-V')+\frac12(X-X')^i\Delta_{ij}(u,u')(X-X')^j\ ,
\label{ab1}
}
where
\EQ{
\Delta_{ij}(u,u')=(u-u')\Big[\int^u_{u'}du''\,C^{-1}(u'')\Big]^{-1}_{ij}\ .
}
The other quantity in \eqref{db} is the Van Vleck-Morette VVM determinant which only depends on $u$ and $u'$ 
in a plane wave geometry:
\EQ{
\Delta(u,u')=\frac1{\sqrt{g(u)g(u')}}\det\Delta_{ij}(u,u')\ .
\label{vvm}
}

Looking at the form of \eqref{ab1}, it becomes clear the $X^{\prime i}$ integrals in \eqref{ab2} are Gaussian, 
while the $V'$ yields a delta function. Before we perform these integrals, it is useful to change variables 
from the proper times $\{T_i\}$, to the set $\{T,\xi_i\}$, where
\EQ{
T_i=T\xi_i\ ,\qquad\sum_{i=1}^{n-1}\xi_i=1\ .
}
where the parameters $\xi_i\in[0,1]$. We also define
\EQ{
\frac1{\zeta}=\sum_i\frac1\xi_i\ .
}
The relevant Jacobian is
\EQ{
\int\prod_{i=1}^{n-1}\frac{dT_i}{T_i^{d/2}}=\int\frac{dT}{T^{(n-1)(d/2-1)+1}}\prod_{i=1}^{n-1}
\frac{d\xi_i}{\xi_i^{d/2}}\,\delta\big(1-\sum_{i=1}^{n-1}\xi_i\big)\ .
}

The $V'$ integral yields a delta function:
\EQ{
\int dV'\,\exp\Big[\frac{(u'-u)V'}{2iT\zeta}-iE V'\Big]=4\pi T\zeta\delta\big(u'-u+2\zeta ET\big)\ ,
\label{ab3}
}
while the transverse coordinates yield
\EQ{
\int d^{d-2}X'\,\exp\Big[\frac i{4T\zeta}\,X^{\prime i}\Delta_{ij}(u,u')X^{\prime j}\Big]&=
\big(4\pi i\zeta T\big)^{(d-2)/2}\big(\det\Delta_{ij}(u,u')\big)^{-1/2}\\
&=\big(4\pi i\zeta T\big)^{(d-2)/2}(gg')^{-1/4}\Delta(u,u')^{-1/2}\ .
}

The delta function in \eqref{ab3}, can be used to exchange the $T$ integral for one over the separation 
$t=u-u'$. Finally, after integrating over $u$,  we have the following expression for the eikonal phase
\EQ{
\Theta(u)=\frac{2^{p-1}i^{1-p}}{(4\pi)^{(n-2)d/2}}\lambda^2E^{p-2}\int d^{n-1}\xi\,\chi(\xi_i)
\int_{-\infty}^udu\,\int_0^{\infty-i0^+}\frac{dt}{t^p}\,e^{-izt}\Delta(u,u-t)^{(n-2)/2}\ ,
\label{bb1}
}
where
\EQ{
z=\frac{m^2}{2E}\big(\zeta^{-1}-1\big)
}
and 
\EQ{
\int d^{n-1}\xi\equiv\int \prod_{i=1}^{n-1}d\xi_i\,\delta\big(1-\sum_i\xi_i\big)\ ,\qquad
\chi(\xi_i)=\zeta^{(n-1)(d-2)/2}\prod_{i=1}^{n-1}\xi_i^{-d/2}\ .
}

The prescription explicit in the definition of the $t$ integral is needed because the VVM determinant 
can have singularities on the real $t$ axis and these must be avoided by moving into the lower half plane.

Before we turn to evaluating the phase, we first note that on physical grounds we must have $d\geq4$ and 
so $p\geq1$. Note also that $p=2$ when $d=d_\text{crit}$. The expression in \eqref{bb1} has divergences
which can arise when $t\to0$, {\it i.e.}~$u\to u'$. These are the usual UV divergences that are expected 
even in flat space. 
When $d>d_\text{crit}$, we will see that there are new curvature-dependent UV divergences, as one might 
have expected in a non-renormalizable theory.

There are also divergences that can arise when $u\to-\infty$. These arise from the way that \eqref{aa1} 
is treated as an initial-value problem. The proper way to formulate the problem is in terms of the 
Schwinger-Keldysh formalism as an initial value problem. In  a plane wave geometry, it is convenient 
to work in the light front formalism and choose an initial value surface as the light front at some finite 
$u=u_0$. The limit $u_0\to-\infty$ may be taken later. One way to choose boundary conditions is 
to suppose that the interaction, which in this example is $\lambda\phi^n$, turns on abruptly at $u=u_0$. 
In that case, the field is free for $u<u_0$ and when the coupling is turned on it becomed dressed in 
real affine time for $u>u_0$. The picture is that a cloud of virtual quanta builds up around the bare state.
This process can have divergences that in the flat space theory are absorbed into wave function 
renormalization. In curved space, there can be additional curvature-dependent divergences that require 
additional wave function renormalization. 

In order to disentangle the flat space and curvature-dependent divergences in \eqref{bb1}, 
we can separate out the curvature-dependent effects we are interested in by subtracting 
the flat space contribution from \eqref{bb1}. This gives\footnote{In this equation 
${\cal C}=2^{p-1}(4\pi)^{-(n-2)d/2}$ is a constant.}
\EQ{
\Theta(u)=\lambda^2{\cal C}i^{1-p}E^{p-2}\int d^{n-1}\xi\,\chi(\xi_i)
\int_{-\infty}^udu\,\int_0^{\infty-i0^+}\frac{dt}{t^p}\,e^{-izt}\big(\Delta(u,u-t)^{(n-2)/2}-1\big)\ .
\label{bbb2}
} 
From now on, we will work with this subtracted phase. Also, it should be pointed out that $\Theta(u)$ 
has both real and imaginary parts. As far as questions of causality are concerned, we are interested 
in the asymptotic value of the real part which gives the Shapiro time delay, or advance, via
\EQ{
\Delta v=\RE\frac{\Theta}E\ ,\qquad \Theta_\text{scat}=\Theta(u\to\infty)\ .
}

\section{Shockwave Geometries}\label{s3}

The Aichelburg-Sexl metric for a gravitational shockwave, in $d$ dimensions,  
is given in \eqref{ba}.\footnote{In terms of Cartesian coordinates, we take $u=\frac12(t+z)$ and 
$v=t-z$.} The profile function $f(r)$ is determined by the Ricci curvature $R_{uu} = 8\pi G T_{uu}$
and depends on the nature of the matter source. 

We will consider two kinds of source: (i) a particle
boosted into a frame where it has a large energy and, effectively, moves along a null geodesic 
$v=x^j=0$ with an energy-momentum tensor $T_{uu} = E'\delta^{n-2}(x^i)\delta(u)$; and (ii) a beam 
corresponding to a uniform energy density  boosted into the same frame so that $T_{uu}=\mu\delta(u)$.

The corresponding profiles $f(r)$ follow from the relation $R_{uu} = - \tfrac{1}{2} \Delta f(r)$, 
where $\Delta$ is the $d-2$-dimensional Laplacian. This gives
\EQ{
\text{(particle)}:&\quad f(r)=\frac{4\Gamma(\frac{d-4}2)}{\pi^{\frac{d-4}2}}\cdot\frac{GE'}{r^{d-4}}\ ,\\
\text{(beam)}:&\quad f(r)=-\frac{8\pi G\mu}{d-2}r^2\ .
\label{bb}
}

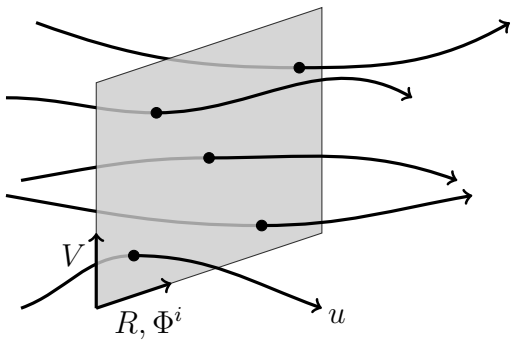
\begin{figure}[h]
\begin{center}
\begin{tikzpicture}[scale=1,decoration={markings,mark=at position 0.5 with {\arrow{>}}}]
\draw[very thick] (0.5,0.7) to[out=180,in=20] (-1,0);
\draw[very thick] (2.2,1.1) to[out=180,in=-10] (-1.2,1.5);
\draw[very thick] (2.7,3.2) to[out=180,in=-20] (-0.8,3.8);
\draw[very thick] (0.8,2.6) to[out=180,in=0] (-1.2,2.8);
\draw[very thick] (1.5,2) to[out=180,in=10] (-1,1.7);
\draw[fill=black!20,opacity=0.8] (0,0) -- (3,1) -- (3,4) -- (0,3) -- (0,0);
\draw[very thick,->] (0,0) -- (1,0.33333);
\draw[very thick,->] (0,0) -- (0,1);
\draw[very thick,->] (0.5,0.7) to[out=0,in=160] (3,0);
\draw[very thick,->] (2.2,1.1) to[out=0,in=-170] (5,1.5);
\draw[very thick,->] (2.7,3.2) to[out=0,in=-150] (5.5,3.8);
\draw[very thick,->] (0.8,2.6) to[out=0,in=150] (4.2,2.8);
\draw[very thick,->] (1.5,2) to[out=0,in=160] (4.8,1.7);
\filldraw[black] (0.5,0.7) circle (2pt);
\filldraw[black] (2.2,1.1) circle (2pt);
\filldraw[black] (2.7,3.2) circle (2pt);
\filldraw[black] (0.8,2.6) circle (2pt);
\filldraw[black] (1.5,2) circle (2pt);
\node at (0.7,-0.2) {$R,\Phi^i$};
\node at (-0.3,0.7) {$V$};
\node at (3.2,-0.1) {$u$};
\end{tikzpicture}
\caption{\footnotesize Adapted coordinates for a congruence of null geodesics. 
The null coordinate $V$ and $d-2$ space-like coordinates $(R,\Phi^i)$ label the individual geodesics 
in the congruence, while the other null coordinate $u$ plays the role of the affine coordinate along 
the geodesics.}
\label{f5}
\end{center}
\end{figure}

The null geodesics corresponding to the trajectories of high energy particles propagating
in the $u$-direction in this background are well-known and display a discontinuous jump in the
Aichelburg-Sexl $v$ coordinate as the particle crosses the shockwave. In polar coordinates 
for the transverse space,
\EQ{
v &= V + \frac{1}{2} f(R) \vartheta(u) + \frac{1}{8} f'(R)^2 u \vartheta(u)  \ ,\\
r &=  R+ \frac{1}{2} f'(R) u \vartheta(u) \ , \\
\phi^i&=\Phi^i   \ .
\label{bc}
}
$i=1,2,\ldots,d-3$ label the angular coordinates. The 
$V,R,\Phi^i$ are constants labelling the individual geodesics in a null congruence: see fig.~\ref{f5}.
They are therefore natural ``adapted coordinates'', in terms of which the Aichelburg-Sexl metric
can be rewritten as
\begin{equation}
ds^2 = - 2 du \,dV + \left[1 + \frac{1}{2}f''(R) u \vartheta(u)\right]^2 dR^2
+\left[1 + \frac{1}{2R} f'(R) u \vartheta(u)\right]^2 R^2 d\Phi^i\,d\Phi^i\ .
\label{bd}
\end{equation}

These geodesics describe straight, null trajectories in both half-planes $u<0$ and $u>0$
with a discontinuous coordinate shift $\Delta v= \tfrac{1}{2} f(R)$ and a deflection angle $\phi$,
with $\tan\phi/2 =  -\tfrac{1}{2}f'(R)$, at $u=0$. The full shockwave spacetime can therefore be viewed as
two Minkowski half-planes patched together along the surface $u=0$ with a displacement $\Delta v$. 

Now, as discussed extensively in our earlier work, the effect of vacuum polarisation on the propagation
of a particle in a curved spacetime background depends on the geometry of geodesic deviation.
This is precisely the feature of the background that is encoded in the Penrose limit \cite{Penrose} 
(see also \cite{Blau:2006ar}).
The Penrose limit is a plane-wave spacetime which is determined from the original spacetime metric 
{\it and} a preferred geodesic. In a general spacetime, in adapted coordinates with preferred geodesic
$V = X^i = 0$, $i = 1,2,\ldots,d-2$, the metric may be written as
\begin{equation}
ds^2 = - 2 du\,dV + C(u,V,X^i) dV^2 + 2 C_i(u,V,X^i) dX^i\, dV + C_{ij}(u,V,X^i) dX^i\, dX^j \ .
\label{be}
\end{equation}
The Penrose limit is then
\EQ{
ds^2\Big|_\text{Penrose} = \lim_{\lambda\rightarrow0} ~~
\frac{1}{\lambda^2}~ ds^2(u,\lambda^2 V, \lambda X^i) = - 2 du \,dV + C_{ij}(u,0,0) dX^i \,dX^j  \ .
\label{bf}
}

For the Aichelburg-Sexl shockwave, we choose a preferred geodesic with impact parameter $b$,
{\it i.e.}~$V=0$, $R=b$, $\Phi^i=0$, so that $X^i = b\Phi^i$, $i=1,2,\ldots,d-3$,  $X^{d-2} = R-b$. 
The Penrose limit is then \cite{Hollowood:2009qz}
\begin{equation}
ds^2\Big|_\text{Penrose} =~- 2 du \,dV + C_{ij}(u) dX^i\, dX^j \ ,
\label{bg}
\end{equation}
where
\EQ{
C_{ij}(u)=\big[1-\sigma_i u\vartheta(u)\big]^2\delta_{ij}\ ,
}
and we have defined
\EQ{
\text{(particle)}:&\quad\sigma_i\equiv\sigma=\frac{4\Gamma(\frac{d-2}2)}{\pi^{\frac{d-4}2}}
\cdot\frac{GE'}{b^{d-2}}\ ,\quad i=1,2,\ldots,d-3\ ,\\ &\quad \sigma_{d-2}=-(d-3)\sigma\ ,\\
\text{(beam)}:&\quad\sigma_i\equiv\sigma=\frac{8\pi G\mu}{d-2}\ ,\quad i=1,2,\ldots,d-2\ .
}
Note that in the particle case that $\sum_{i=1}^{d-2}\sigma_i=0$ which means that $R_{uu}=0$, 
{\it i.e.}~the geometry is Ricci flat. This is to be expected, since the geometry is a vacuum solution 
everywhere except at the position of the particle.

This is written in Rosen coordinates, which are well-suited to describing the geodesic congruence.
An alternative presentation is in terms of Brinkmann coordinates, where the metric is instantly
recognisable as a plane wave:
\EQ{
ds^2\Big|_\text{Penrose}=-2du\,dv  +\sum_{i=1}^{d-2}\Big[\sigma_ix^ix^i\delta(u)du^2 +dx^i\,dx^i \Big]\ .
\label{bi}
}
We will not have use for these coordinates in this paper.

The VVM determinant for this geometry is, from \eqref{vvm}, 
\EQ{
\Delta(u,u')=\prod_{i=1}^{d-2}\frac{|u-u'|}{|u-u'|+\sigma_iuu'}\ ,\quad uu'<0\ ,\qquad\Delta(u,u')=1\ ,
\quad uu'>0\ .
}
Note that $\Delta(u,u')$ is only non-trivial if $u$ and $u'$ lie on opposite sides of the plane of the shockwave.

\section{Analysing the Eikonal Phase}

In this section, we will investigate the eikonal phase for a shockwave spacetime. 
The goal is to evaluate the eikonal phase $\Theta(u)$ in the asymptotic limit where $u\to\infty$,
since this is the quantity relevant for scattering. 
In general, the $u$-dependence of $\Theta(u)$ exhibits interesting behaviour in its own right,
as explored in \cite{Hollowood:2015elj}, especially near the focal point of the null geodesic congruence.
Here, however, we are just interested in the scattering phase $\Theta_{\rm scat}(\hat{s}) = 
\Theta(u\rightarrow\infty)$ and especially in its behaviour when the energy $E$ is small and large. 

If we use the dimensionless coupling $\tilde\lambda$, the eikonal phase then depends only on the 
dimensionless quantity $\hat{s}$ introduced above:
\EQ{
\Theta_\text{scat}=\tilde\lambda^2\,{\mathscr F}(\hat s)\ ,\qquad\hat s=\frac{\sigma E}{m^2}\ ,
}
so the regimes of low and high energy are more precisely defined in terms of $\hat{s}$,
\EQ{
\text{(Low energy):}\quad\hat s\ll1\ ,\qquad\text{(High energy):}\quad\hat s\gg1\ .
}

The expression for the subtracted phase is \eqref{bbb2} and so we simply have to take the 
limit $u\to\infty$. 
Note that for the shockwaves that $\Delta(u,u')=1$ when $u$ and $u'$ lie on opposite sides 
of the shockwave. This means that in \eqref{bbb2} the $t$ integral can be taken to have a lower limit 
$u$ and the lower limit of the $u$ limit can be taken to $0$:
\EQ{
\Theta_\text{scat}=\lambda^2{\cal C}i^{1-p}E^{p-2}\int d^{n-1}\xi\,\chi(\xi_i)
\int_0^\infty du\,\int_u^{\infty-i0^+}\frac{dt}{t^p}\,e^{-izt}\big(\Delta(u,u-t)^{(n-2)/2}-1\big)\ .
\label{bb3}
}

It is useful to reverse the order of the $t$ and $u$ integrals, using
\EQ{
\int_0^\infty du\int_u^{\infty-i0^+}dt=\int_0^{\infty-i0^+}dt\int_0^tdu\ .
}
Since the shockwaves just depend on one curvature scale $\sigma$, it makes sense to scale this 
out of the integrals by taking $t\to t/\sigma$ and $u\to u/\sigma$. In addition, to take account 
of the prescription on the $t$ integral, we rotate the contour $t\to-it$. Taking all this into account leaves
\EQ{
\Theta_\text{scat}=\tilde\lambda^2{\cal C}\hat s^{p-2}\int d^{n-1}\xi\,\chi(\xi_i)\,\int_0^{\infty} 
\frac{dt}{t^p}\,e^{-\hat zt}J(t)\ ,
\label{bb2}
}
where $\hat z=z/\sigma=(\zeta^{-1}-1)/(2\hat s)$ is dimensionless.

In \eqref{bb2}, we have defined
\EQ{
\text{(particle):}&\quad J(t)=\int_0^tdu\,\Big[t^p\big(t+iu(u+it)\big)^{-p_1}
\big(t-i(d-3)\sigma u(u+it)\big)^{-p_2}-1\Big]\ ,\\[7pt]
\text{(beam):}&\quad J(t)=\int_0^tdu\,\Big[t^p\big(t+iu(u+it)\big)^{-p}-1\Big]\ ,
\label{bb3}
}
where $p_1=(n-2)(d-3)/2$, $p_2=(n-2)/2$ with $p_1+p_2=p$. 

In the beam case, $J(t)$ can be evaluated for arbitrary $p$,
\EQ{
J(t)=&-\frac{2^{-p-1} it}{p-1}(y^{-1}-y)\Big\{(1+y)^{p-1}\, _2F_1\left(1-p,p,2-p;
\tfrac12(1-y)\right)\\ &-
(1-y)^{p-1}\, _2F_1\left(1-p,p,2-p;\tfrac12(1+y)\right)\Big\}+it\ ,
\label{uur}
}
where $y=(1+4i/t)^{-1/2}$. For the particle case, $J(t)$ can be evaluated for particular $p$ but the 
expressions are cumbersome and we will not write them here.

In \cite{Hollowood:2015elj}, we presented a number of numerical plots of the corresponding results
for QED to illustrate the full $\hat{s}$-dependence of the phase and how it interpolates between the
low-energy and UV limits. Here, we are most interested in the limiting behaviours and their relevance
for causality, so we present only these results below.

\subsection{Low energy expansion}

The expansion in the energy $E$, or equivalently the curvature, follows from expanding $J(t)$ 
in powers of $t$:
\EQ{
\text{(particle):}&\quad J(t)=i\frac{(n-2)(d-2)(d-3)}{120}t^3-\frac{(n-2)(d-2)(d-3)(d-4)}{840}t^4+
\cdots\ ,\\[7pt] \text{(beam)}:&\quad J(t)=-\frac{p}{6}t^2+i\frac{(p+1)p}{60}t^3+
\frac{(p+2)(p+1)p}{840}t^4+\cdots\ .
}

We can then perform the $t$ integral on each of the terms above separately, using
\EQ{
\int_0^{\infty}dt\,e^{-\hat z t}t^a=\hat z^{-a}\Gamma(1+a)\ .
}
The expansion of the eikonal phase consequently takes the form
\EQ{
\Theta_\text{scat}=-b_1\Gamma(3-p)\tilde\lambda^2\hat s+ib_2\Gamma(4-p)\tilde\lambda^2\hat s^2-
b_3\Gamma(5-p)\tilde\lambda^2\hat s^3+\cdots\ .
\label{hh1}
}
The coefficients $b_i$ are positive real numbers depending on $d$ and $n$. The gamma functions 
in this formula encode the UV structure of the theory. Divergences appear when $p>2$, 
in other words when $d>d_\text{crit}$. 
So when the theory becomes  non-renormalizable there are additional curvature dependent divergences. 
These can be regularized with additional curvature-dependent counterterms.

We recognize the first term in \eqref{hh1} as $E\Delta v$, with the Shapiro time advance 
in \eqref{yy1}. This term is absent for the particle shockwave because that geometry is a 
vacuum solution and so is Ricci flat. We have therefore recovered the effective action prediction
as the low-energy limit of our general result.

\subsection{High energy expansion}

The large $E$ behaviour is encoded in the large $t$ behaviour of $J(t)$. This is ${\mathscr O}(t^{1-p})$, 
therefore when $p>2$, that is $d>d_\text{crit}$, the $t$ integral is convergent at its upper limit when 
$\hat z=0$. The lower limit $t\to 0$ is divergent. These are the UV divergences already identified. 
In order to make these completely explicit, we will introduce 
an explicit high-momentum Euclidean cut off $\Lambda$. This appears as a cut-off on the lower limit 
of the $T$ integral of $-i\Lambda^{-2}$ which becomes a cut-off on the lower limit of the $t$ integral 
of $2i\zeta E/\Lambda^2$. After the re-scaling and analytic continuation, this becomes 
$\delta=2\zeta\sigma E/\Lambda^2$.

So when $d>d_\text{crit}$, that is $p>2$, we have 
\EQ{
\int_\delta^\infty \frac{dt}{t^p}\,J(t)=\text{const.}+\sum_{j=1}^{[p-2]}c_j\delta^{p-2-j}\ ,
}
where the $j=p-2$ term, if present, is $\delta^0\to\log\delta$.
So assuming that the UV divergences are cancelled by counterterms, the high energy behaviour is simply
\EQ{
(p>2):\qquad\Theta_\text{scat}=\tilde\lambda^2\tilde{\cal C}\hat s^{p-2}+{\mathscr O}(\hat s^{p-3})\ . 
\label{gtt1}
}
for a complex constant $\tilde{\cal C}$.

When $d\leq d_\text{crit}$, that is $p\leq2$, there are no UV divergences. However, the $t$ integral 
is no longer convergent at its upper end when $z=0$. This means that the behaviour is richer than 
$E^{p-2}$. We find   
\EQ{
(p=2):\qquad&\Theta=\tilde\lambda^2\big(-\tilde{\cal C}_1+
i\tilde{\cal C}_2\log\hat s+{\mathscr O}(\hat s^{-1}\log\hat s)\big)\ ,\\[3pt]
(p=\tfrac32):\qquad&
\Theta=\tilde\lambda^2\big(-\tilde{\cal C}_1\hat s^{-1/2}+i\tilde{\cal C}_2+
{\mathscr O}(\hat s^{-3/2})\big)\ ,\\[3pt]
(p=1):\qquad&\Theta=\tilde\lambda^2\big(-\tilde{\cal C}_1\hat s^{-1}\log^2\hat s+
i\tilde{\cal C}_2+{\mathscr O}(\hat s^{-2}\log^2\hat s)\big)\ .
\label{gtt2}
}
where the $\tilde{\cal C}_i$ are real positive constants.

The high energy behaviour \eqref{gtt1} and \eqref{gtt2} of the eikonal phase is our main result. 
Physically, we have $d\geq4$, so the results apply to a variety of $\phi^n$ theories in different 
dimensions:
$\phi^3$ theories in $d=4$ ($p=1$), $d=5$ ($p=3/2$), $d=6$ ($p=2$) and $d>6$ ($p>2$); 
$\phi^4$ in $d=4$ ($p=2$), and $d>4$ ($p>2$); 
and $\phi^n$ for $n>4$ in $d>4$ ($p>2$).

It is then apparent that when the theory is (perturbatively) renormalizable, $p\leq2$, 
the eikonal phase remains perturbatively bounded. So in this case the theory has a good UV completion 
and in the high energy regime the Shapiro time advance goes to zero:
\EQ{
(p\leq2):\qquad\Delta v(E\to\infty)=0\ .
}
The Shapiro time advance induced by the Ricci term in the effective action \eqref{cf1} therefore does
{\it not} lead to causality paradoxes.

On the other hand, above the critical dimension, $p>2$ and the eikonal phase grows at high energy. 
The causality problems are therefore not resolved and at some point perturbation theory breaks down.
This is an indication that the lack of a well-defined UV completion implies that the causal problems inherent  
in the low energy effective action are not resolved at high energy.

\section{Summary and Outlook}

In this paper, we have studied ultra-high energy scattering in scalar QFTs with different renormalizability properties 
in order to shed further light on the relation of causality to the existence of a UV completion and how the
apparent causal paradoxes arising in the associated effective theories are resolved.
As is well-known, the leading eikonal approximation for the scattering amplitude ${\cal A}(s,t)$ at Planck energies
may be found by studying the propagation of a null particle in an Aichelburg-Sexl shockwave background.
At tree level in the QFT, the particle experiences a discontinuous lightcone coordinate shift as it passes 
through the shockwave wavefront, and the corresponding phase shift $\Theta_{\rm cl}(s,b)$ gives rise to 
the familiar leading-order approximation (\ref{a2}) to ${\cal A}(s,t)$. 

We may go beyond this by considering vacuum polarization loop corrections in the QFT. These introduce a 
new scale $\lambda_c$ characterising the size of the virtual cloud on which the gravitational tidal forces act.
At low energies, below the Planck scale, these can be described by an effective action, which violates the 
strong equivalence principle and in general exhibits superluminal causality violations and Shapiro time 
advances. 

In our recent paper \cite{Hollowood:2015elj}, we showed how this issue is resolved in shockwave backgrounds
by explicitly computing the loop corrections $\Theta_{\rm scat}(\hat{s})$ to the scattering phase for all energies.
As we have seen, this depends only on the single variable 
$\hat{s} = Gs/m^2 b^2 = \left(s/M_p^2\right)\left(\lambda_c/b\right)^2$ 
(in four dimensions) combining the energy and impact parameter.

This calculation is made tractable by the fact that we can use the Penrose limit of the full shockwave 
background in order to perform the loop calculations.  This is because the Penrose limit encodes the geometry 
of geodesic deviation in a tubular neighbourhood of the null trajectory traced out at tree level by the scattered
particle. This technology allows us to extend the evaluation of the scattering phase and coordinate shifts
beyond the effective theory approximation and into the UV limit necessary to address questions of causality.

In this paper, which complements \cite{Hollowood:2015elj}, we have extended these calculations to compute 
the scattering phase shifts $\Theta_{\rm scat}(\hat{s})$ in a variety of self-interacting scalar QFTs with different
renormalizability properties. This allows us to investigate in detail the relationship between causality and 
the existence of a well-defined UV completion.

Our main conclusions are summarized in (\ref{gtt1}) and (\ref{gtt2}).
For the strictly renormalizable theories, we find the perhaps initially surprising result that the scattering phase
does not go to zero in the high-energy limit, but goes asymptotically to a negative constant.
This is, however, still compatible with causality since the corresponding coordinate shift 
$\Delta v$ given by (\ref{zz1}) does indeed vanish in this limit. This precludes the possibility of 
constructing a time machine using the shockwave geometry and causality is respected.
For super-renormalizable theories, the real part of the phase vanishes in the UV limit, with the high
$\hat{s}$ behaviour of $\Theta_{\rm scat}(\hat{s})$ depending on the power-counting parameter $p$
as shown in (\ref{gtt2}).
Finally, for non-renormalizable theories above the critical dimension, we find the phase 
$\Theta_{\rm scat}(\hat{s})$ goes like a positive power of $\hat{s}$ in the UV limit. This implies a
non-vanishing Shapiro time advance even at high energy and is a true violation of causality.

This establishes quite explicitly the expected link between good causal behaviour of the 
scattering amplitude and the existence of a well-defined UV completion of the QFT. 
Note also that the imaginary parts of the phases quoted in (\ref{gtt2}) contribute directly
to the magnitude $\left|{\cal A}(s,t)\right|^2$ of the scattering amplitude.

These results for $\Theta_{\rm scat}(\hat{s})$ determine the scattering amplitude ${\cal A}(s,t)$
from (\ref{a1}) and (\ref{a3}). It would be interesting to investigate further exactly how the energy and
impact parameter dependence we have found here for the phase as a consequence of QFT loop effects
is reflected in the causality and unitarity properties of ${\cal A}(s,t)$ itself.

Finally, it would be interesting to compare these results in more detail with graviton scattering
in string theory, where the string scale $\lambda_s$ plays a r\^ole apparently analogous to that 
of $\lambda_c$ in the QFT case. Here, the reggeization of the scattering amplitude characteristic
of string theory provides the necessary causal UV completion for graviton scattering, where the 
analogous effective action also exhibits causality violation \cite{Camanho:2014apa,D'Appollonio:2015gpa}.

\section*{Acknowledgements}

\noindent This research was supported in part by STFC grant
ST/L000369/1.  We are grateful to Gabriele Veneziano for helpful
discussions and correspondence and the TH Division, CERN for hospitality
during key stages of this work.

\end{document}